**Production and Manufacturing of 3D Printed Acoustic Guitars**

Timothy Tran

Watson School of Engineering, Binghamton University

Professor Schiesser

August 16, 2025

**Abstract**

This research investigates the feasibility of producing affordable, functional acoustic guitars using 3D printing, with a focus on producing structural designs with proper tonal performance. Conducted in collaboration with William Schiesser, the study uses a classical guitar model, chosen for its lower string tension, to evaluate the tonal characteristics of a 3D-printed prototype made from polylactic acid (PLA). Due to the build plate size constraints of the Prusa Mark 4 printer, the guitar body was divided into multiple sections joined with press-fit tolerances and minimal cyanoacrylate adhesive. CAD modeling in Fusion 360 ensured dimensional accuracy in press-fit connections and the overall assembly. Following assembly, the guitar was strung with nylon strings and tested using Audacity software to compare recorded frequencies and notes with standard reference values. Results showed large deviations in lower string frequencies, likely caused by the material choice utilized in printing. Accurate pitches were reached with all strings despite frequency differences through tuning, demonstrating that PLA and modern manufacturing methods can produce affordable, playable acoustic guitars despite inevitable challenges. Further research may investigate alternative plastics for superior frequency matching. This approach holds significant potential for expanding access to quality instruments while reducing reliance on endangered tonewoods, thereby encouraging both sustainable instrument production and increased musical participation. This also creates opportunities for disadvantaged communities where access to musical instruments remains a challenge.

## Production and Manufacturing of 3D Printed Acoustic Guitars

Musical instrument construction has evolved significantly, with guitars progressing from classical designs to modern electric models. While these advancements have expanded musical possibilities, high-quality guitars remain costly, limiting accessibility. This research, conducted with William Schiesser, investigates affordable alternatives by examining the tonal characteristics of 3D-printed acoustic guitars to determine their viability as a viable alternative. By integrating principles of sound design, material science, and manufacturing, the project seeks to identify relationships between material composition, structural design, and sound production. The objective of the present research is to demonstrate how non-traditional materials and modern fabrication techniques can be utilized in musical instrument construction. A greater understanding of 3D printing instruments allows for the production of quality instruments at a lower cost, increasing accessibility for a broader range of musicians.

## 3D Printing Technology and Background

3D printing enables rapid transformation of CAD designs into physical objects, making it a valuable tool for prototyping and manufacturing. In musical applications, 3D-printed electric guitars such as Mikolas Zuza's Prusacaster and Olaf Diegel's printed Gibson demonstrate that printed instruments can work in tandem with electronics.

**Figure 1**

*Mikolas Zuza Prusacaster Design Model*

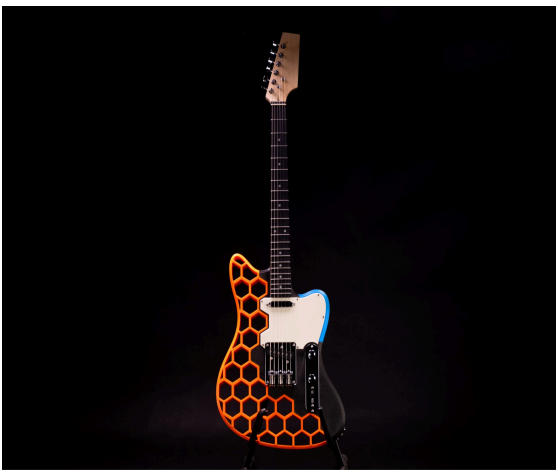

Figure 1 above shows the printed prusacaster, featuring a 3D printed guitar body working in combination with a wooden neck and electronics.

**Figure 2**

*Olaf Diegel Gibson Design Model*

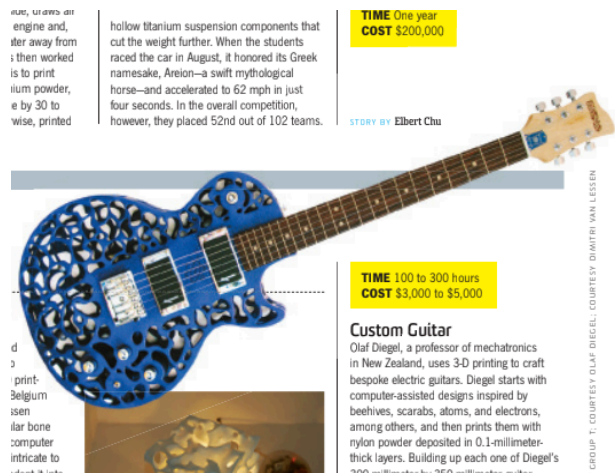

Figure 2 above shows the Olaf Diegel model, also having a fully printed body. Both guitars are examples of 3D printers working alongside musical electronics.

Despite this success, differences from traditional guitar making, aka luthiery, presents challenges in acoustic guitar development. A key drawback would be the difference in design between conventional methods and 3D printing. Effects such as "...the repercussion of the change in printing direction on the acoustic behavior of the different parts printed in PLA…" or changes in the "...concentration of vibration in the bridge area" (Burgos-Pintos, 2023) have been observed in 3D printed guitar front plates. For this project, the primary constraints are printer size and material effects on tonality through string frequency. Polylactic acid, also known as PLA, was selected due to its accessibility and prior use in Pintos's research. This guitar body, unlike traditional guitars, was divided into sections due to the printer's build plate limitations. Press fit tolerancing and adhesives were used to join the guitar body together. Attached below in Table 1 are the reference frequencies for the guitar, sourced from Classical Guitar 101 (Nelson, 2014, para. 5).

**Table 1**

*Classical Guitar String Frequencies*

| 1st String - E4 | (frequency = 329.63) |
|---|---|
| 2nd String - B3 | (frequency = 246.94) |
| 3rd String - G3 | (frequency = 196.00) |
| 4th String - D3 | (frequency = 146.83) |
| 5th String - A2 | (frequency = 110.00) |
| 6th String - E2 | (frequency = 82.41) |

The strings are numbered based on the first string being the highest pitch, which would be the rightmost string on the guitar. Figure 3 depicts the numbered string terminology as shown below.

**Figure 3**

*Classical Guitar String Layout*

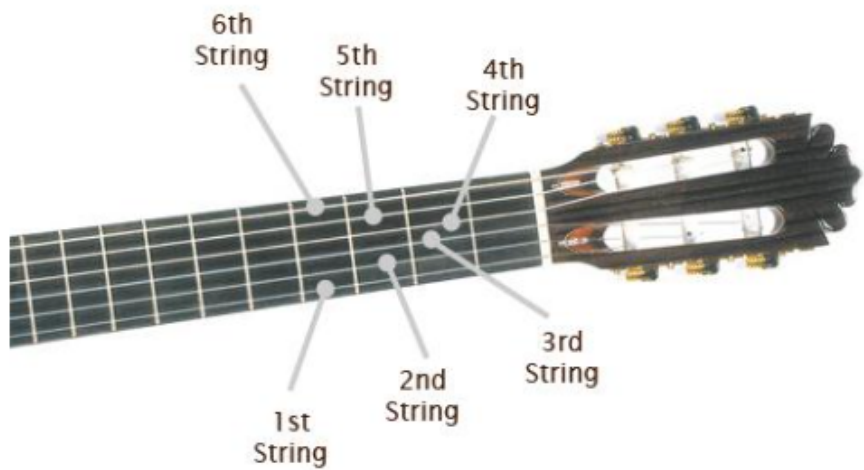

Figure 3 also shows the tuning head and upper fretboard with the tuners and strings installed.

A classical acoustic guitar was chosen as the target model, given that nylon strings have a lower string tension, thus reducing the total tension that the plastic printed body will have to withstand. An older classical guitar was measured and modeled in Fusion 360 as shown in Figure 4, with its main components, including the back plate, front plate, fretboard, and tuning head.

**Figure 4**

*Classical Acoustic Guitar Base Model*

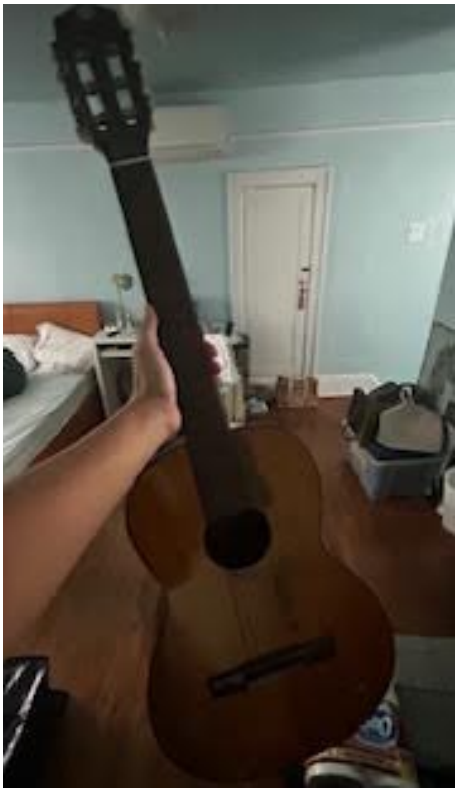

This classical acoustic guitar provides a base-level design, which was measured using a ruler and calipers for linear measurements, and a protractor for angles.

A similar approach to 3D-printed acoustic instruments, as shown in Figure 5 below, has been implemented by Jose Gonzalez, who performs under the name LeFiddler, and uses a 3D-printed violin to deliver performances at weddings and other events (Gonzalez, n.d.). While Gonzalez has integrated some electronics to assist with audio projection, the overall concept is much the same as that used with a traditional violin. Slits in the body are utilized to boost acoustical sound production from the strings, demonstrating that a printed acoustic instrument is certainly a viable and practical method of instrument production that can be adopted for use in a commercial environment.

**Figure 5**

*3D Printed Violin Model used by Gonzalez*

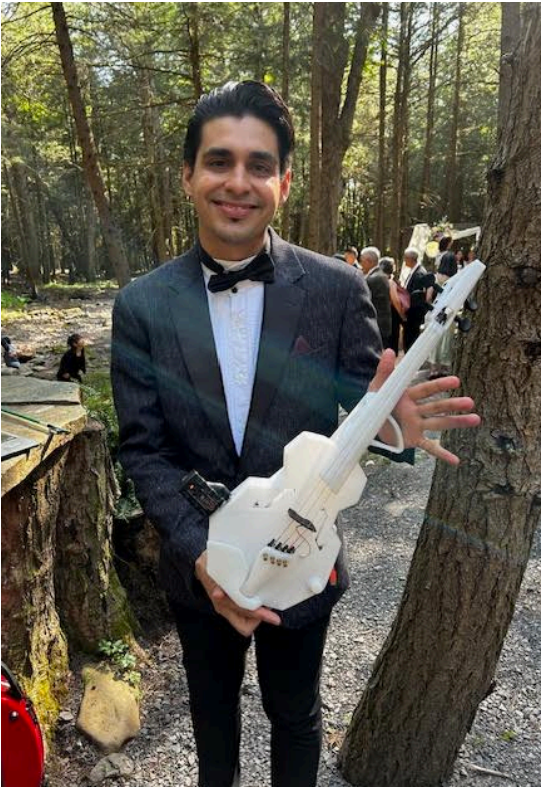

Figure 5 above shows the usage of 3D printed instruments in real life.

<div align="center">

**Support**

</div>

The guitar CAD prototype was initially designed in Fusion 360 based on the guitar shown in Figure 4. The following technical drawings are derived from said prototype, which references the size and shape of the guitar.  The CAD prototype can then be disassembled into modular components that may be created using a 3D printer. The prototype dimensions are measured in inches and are drafted at scales relative to the drawing size. The purpose of each part is described below in its respective figure.

**Figure 6**

*Back Plate Technical Drawing*

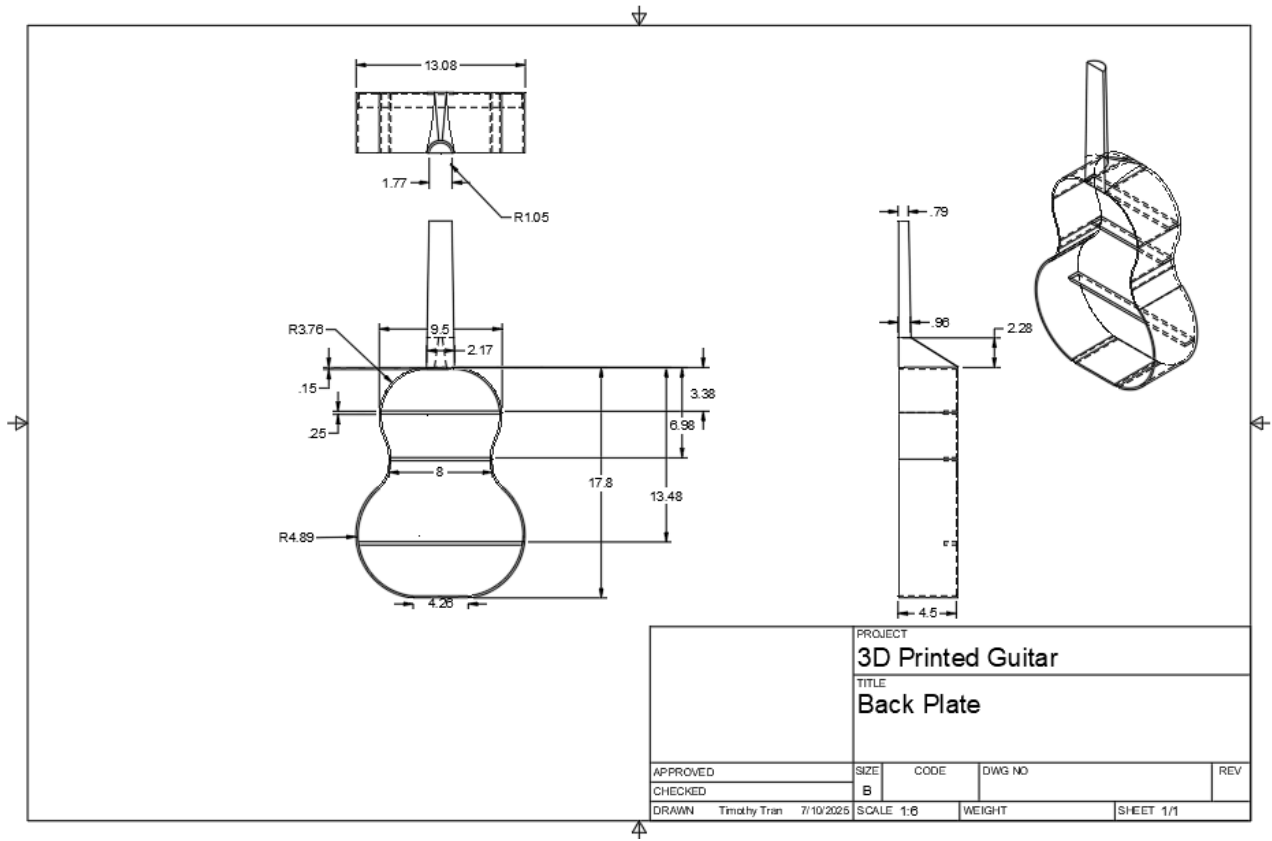

Figure 6 includes the dimensions of the back plate, which comprises the guitar plate, bracing lines, and neck. The bracing lines ensure the structural integrity of the back plate. The neck supports the strings, which are under tension when the guitar is tuned.

**Figure 7**

*Front Plate Technical Drawing*

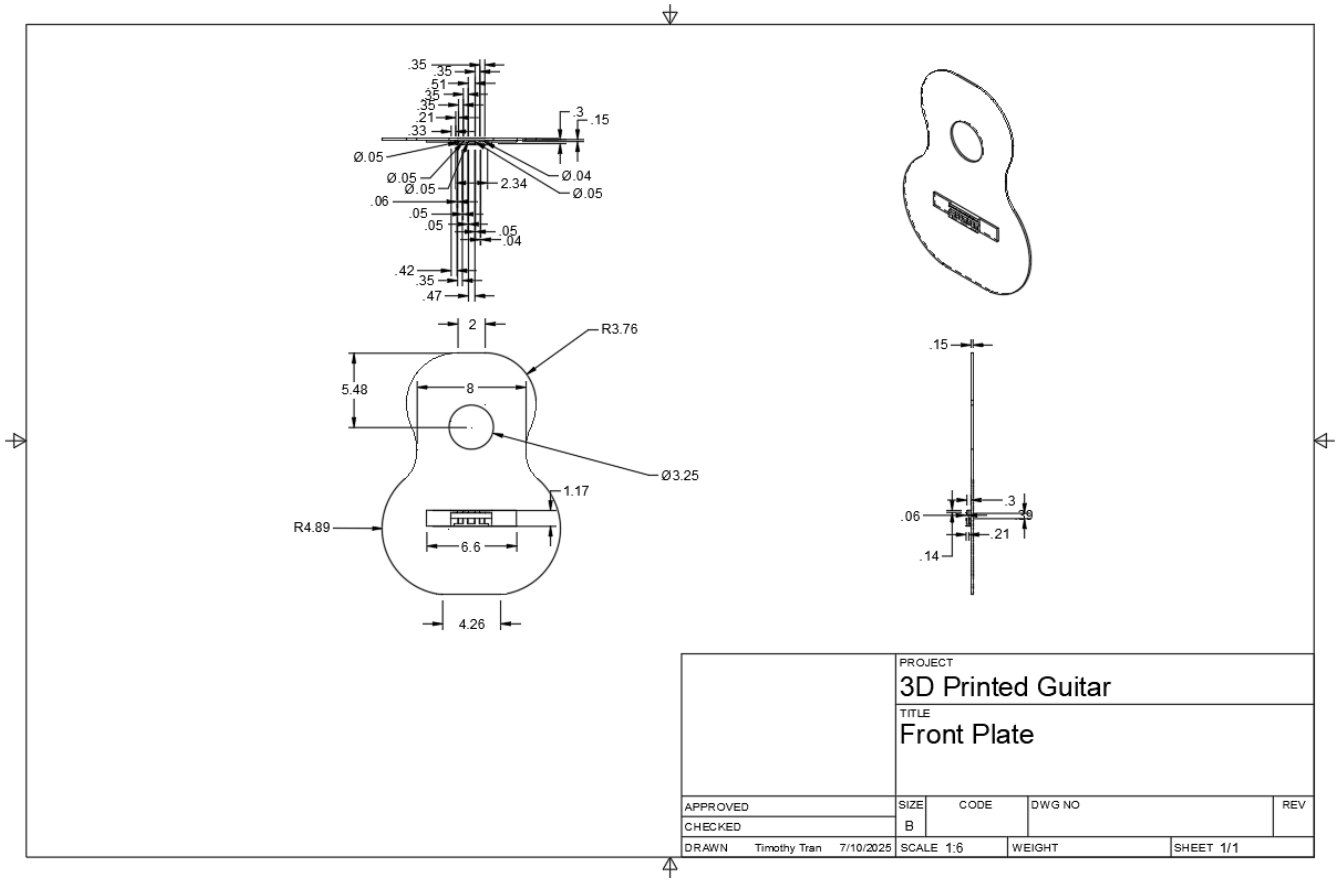

As shown in Figure 7, the guitar front plate comprises the sound hole, the plate itself, the saddle, and the bridge. The saddle and bridge must endure constant string tension while the strings stay in tune.

**Figure 8**

*Fretboard Technical Drawing*

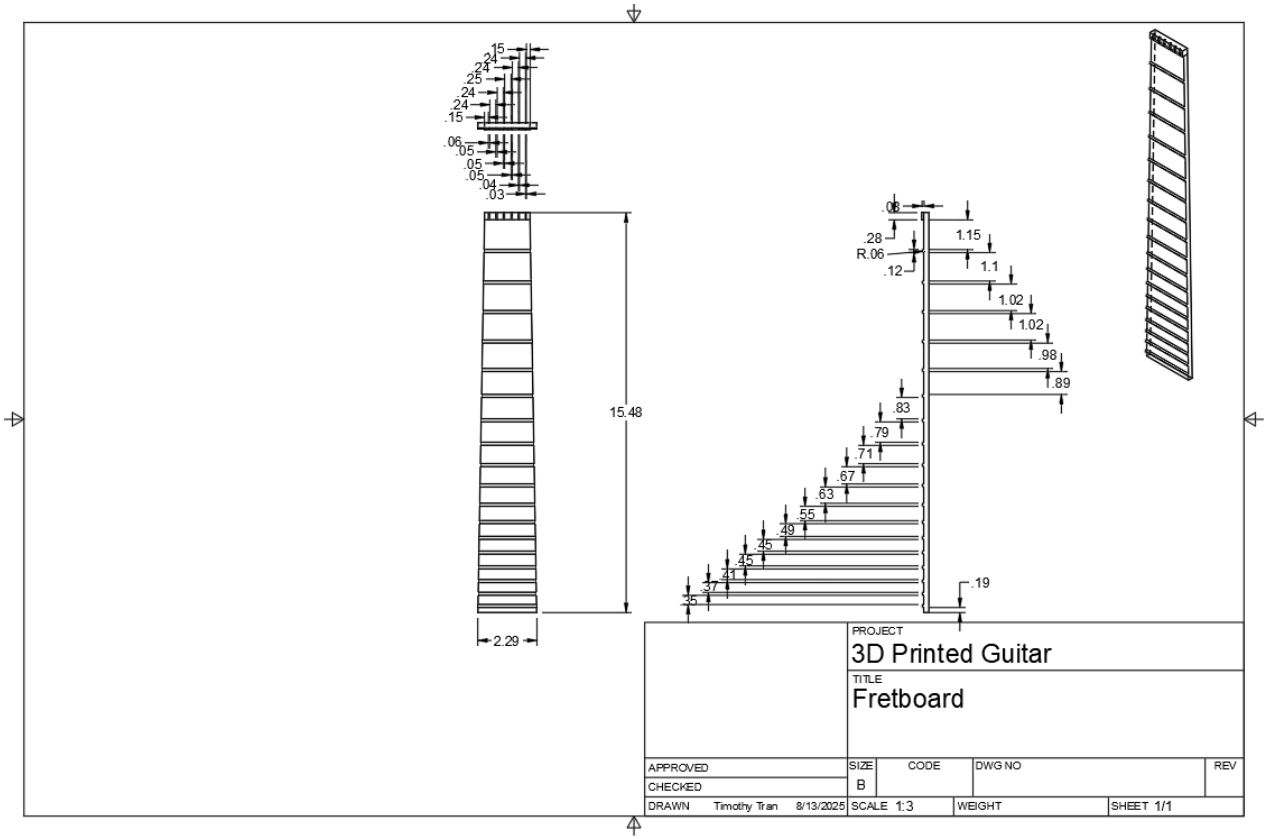

Figure 8 depicts the fretboard, its dimensions, and fret spacing. The distance between each fret decreases toward the bottom of the fretboard to generate higher-pitched notes. The frets must lie flat to create accurate musical notes.

**Figure 9**

*Tuning Head Technical Drawing*

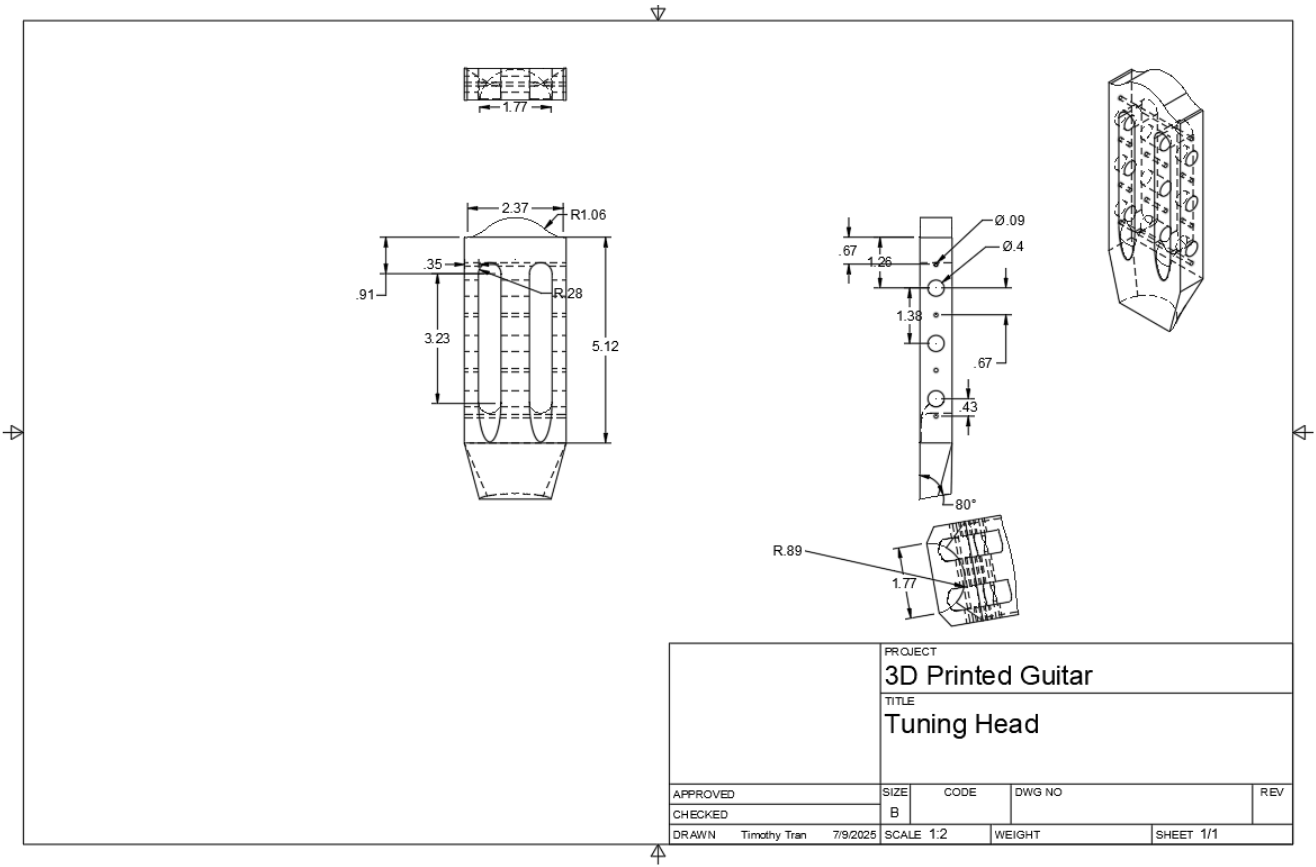

The tuning head is depicted in Figure 9 and is where the guitar string tuners and strings are placed. The tuners hold the strings taut, creating tension between themselves and the bridge.

**Figure 10**

*Base Guitar Model Assembly*

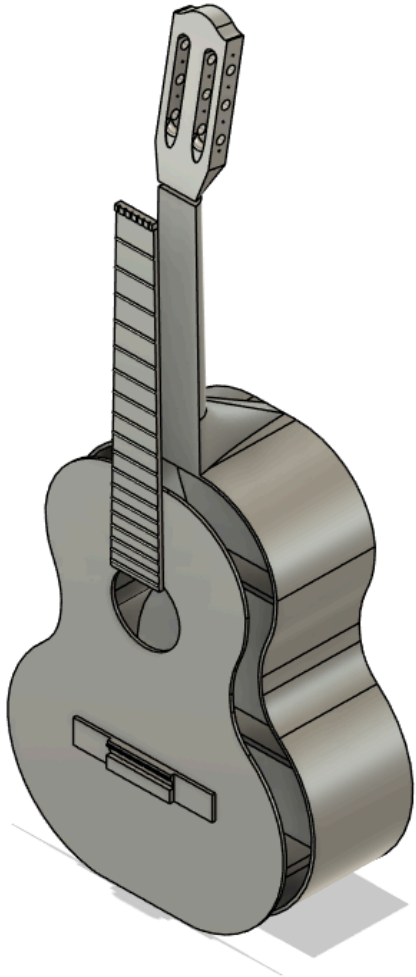

Figure 10 illustrates the complete guitar model assembly. Generic classical guitar tuners and nylon strings are used. Dividing the pieces to fit on the printbed brings into consideration a focus on the adhesion of each piece to construct a functioning guitar.

### Guitar Body Deconstruction Methodology

The Prusa Mark 4 printer's build plate dimensions of 10 × 9.5 inches require a division of the guitar into modular sections. The most practical design utilizes press-fit tolerancing, which involves male and female heads that join each part. Table 2 shows the potential clearances that can be used, as described by Hussain (2019, para. 4).

**Table 2**

*Tolerance Fit Clearance Table*

| Desired Fit | Clearance Gap (in) | Clearance Gap (mm) |
|---|---|---|
| Press Fit | Line to Line | Line to Line |
| Tight Fit | 0.005 | 0.127 |
| Normal Fit | 0.010 | 0.254 |
| Loose Fit | 0.020 | 0.508 |

The clearance gap for a tight fit has a tolerance between .005 and .01 inches.  Thus, .006 inches is used to ensure a tight but adjustable fit. Initial attempts at printing failed due to excessively tight entry points, which prevented the male head from fitting into the female head without cracking. Fillets were then introduced to each female head to ensure smooth assembly. A standard 1 x 1 x 1 inch male head and a 1.00 x 1.01 x 1.01 female head were then explored using test cubes, such as those shown in Figure 11, to great success. Upon test completion, a standard fillet size of 0.03 inches was then used to round the edges with a total tolerance of 0.006 inches.

**Figure 11**

*Test Cube Model*

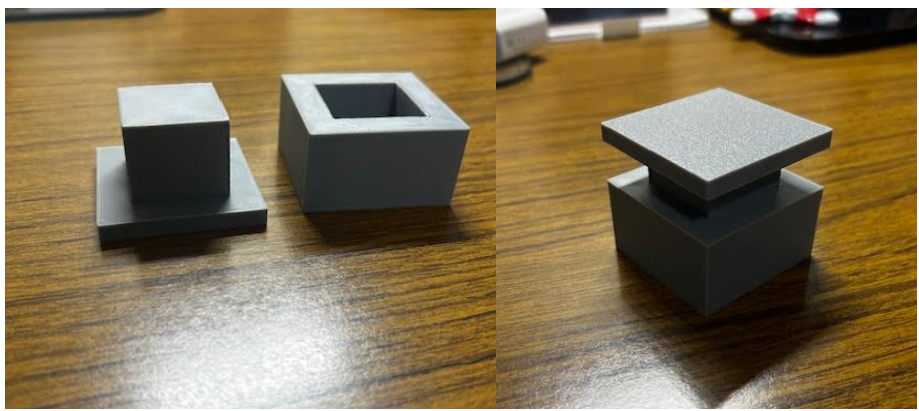

When considering the deconstruction of the guitar model, the resulting division into its individual components must comprise pieces that fit onto the build plate. The following figures provide a global view of the deconstruction of the guitar model into its components.

**Figure 12**

*Deconstructed Back Plate Technical Drawing*

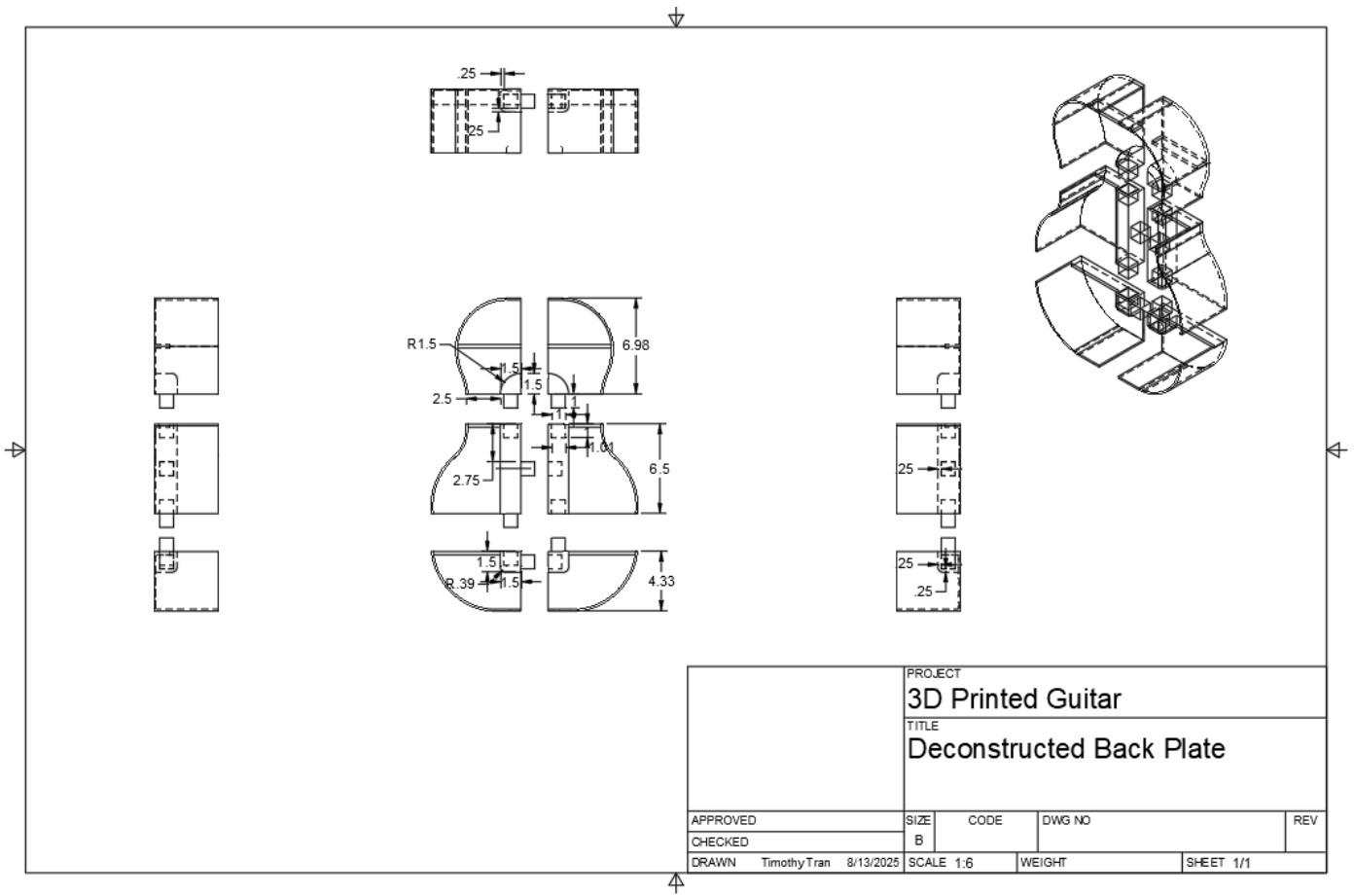

The back plate is divided into six components, implemented via bisecting the body and creating a split along each brace, as shown in Figure 12. Each component includes male and female connection points.

**Figure 13**

*Deconstructed Tuning Head and Neck Technical Drawing*

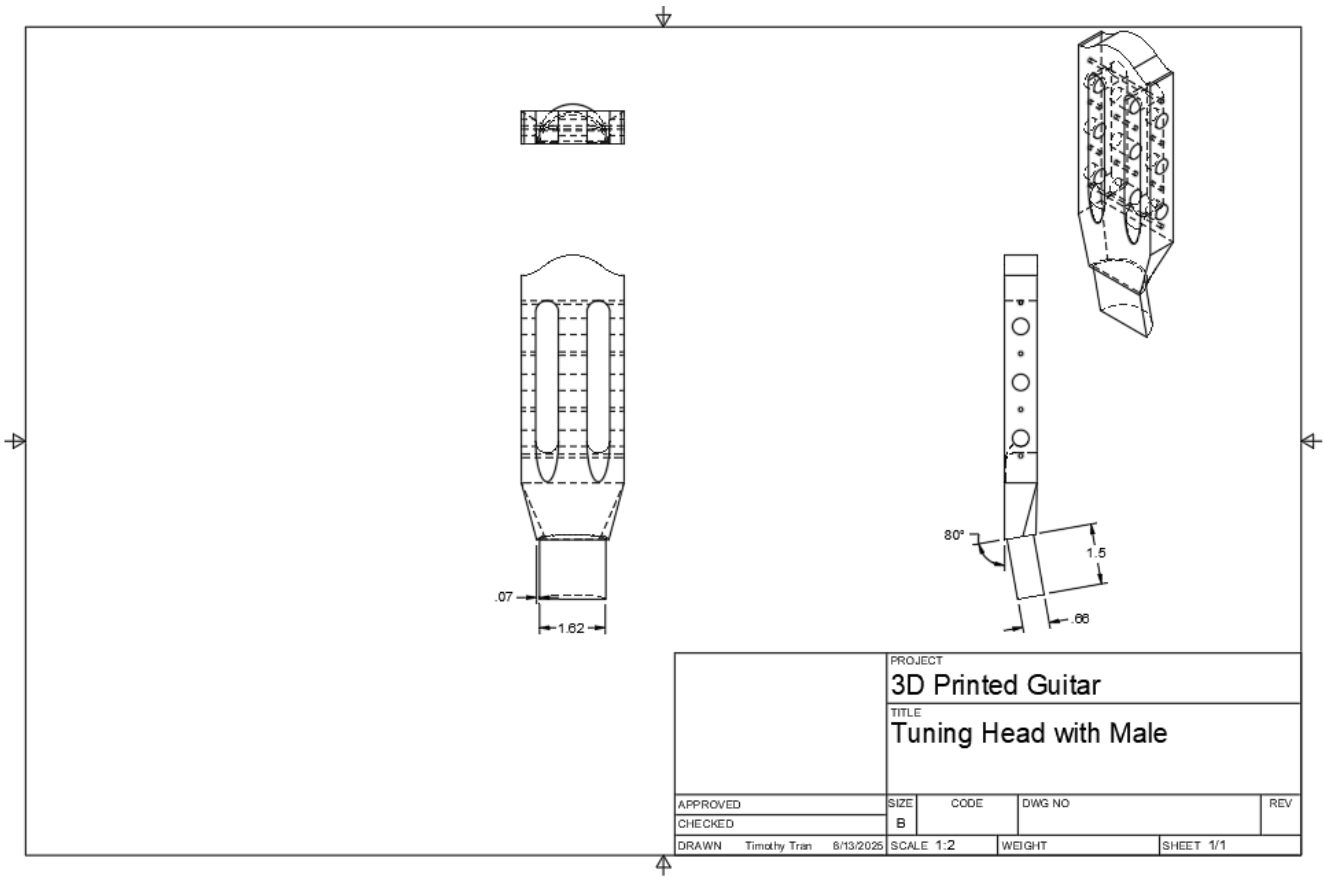

As shown in Figure 13, a male connection point is extended from the tuning head for insertion into the neckpiece. The base of the tuning head is at an angle that promotes proper string flexion when tuned.

**Figure 14**

*Deconstructed Neck Technical Drawing*

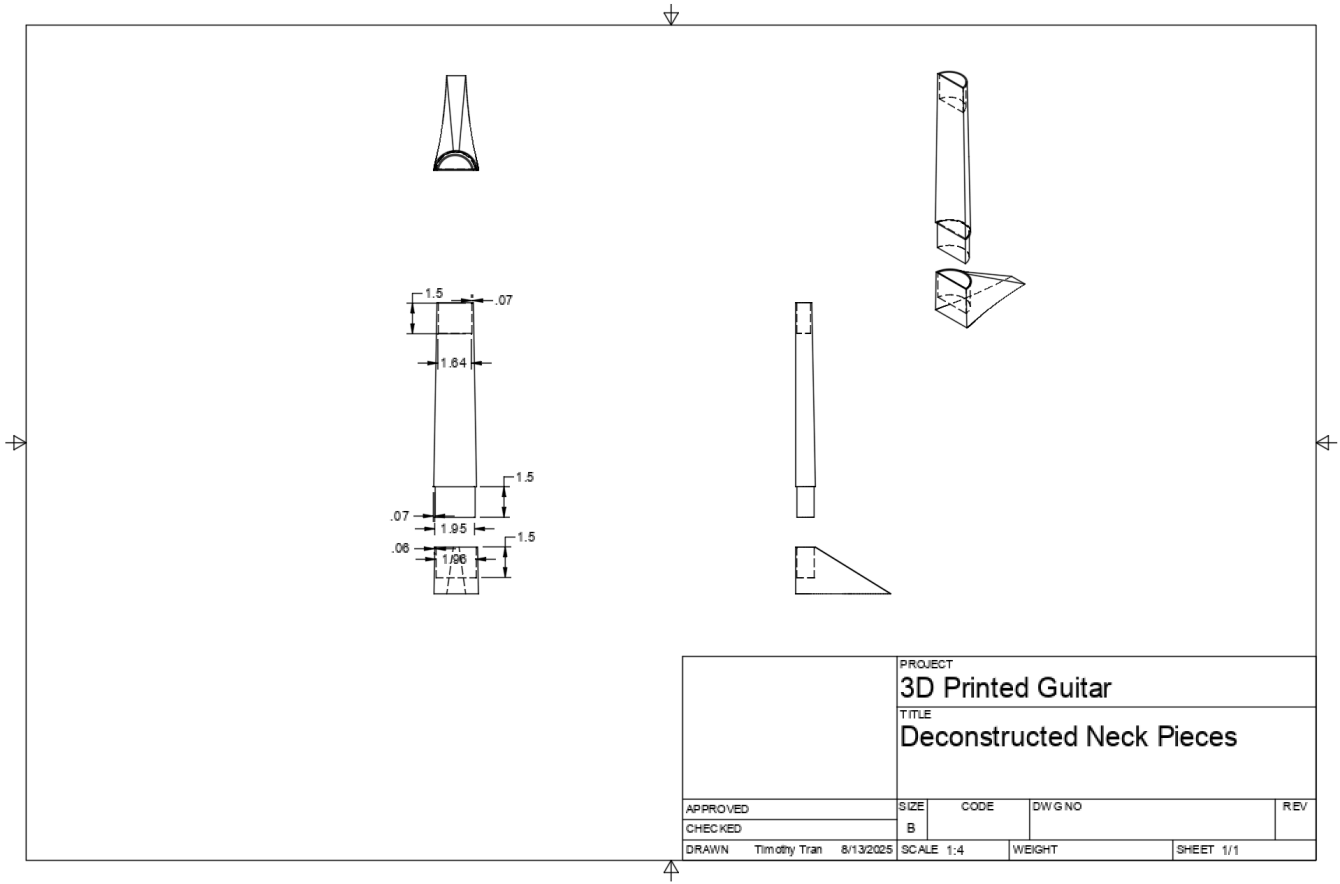

As seen in Figure 14, the neck is detached from the back plate and split into two parts, with the middle neck piece connecting with the tuning head. The initial tuning head design cracked the neck at a clearance of 0.006 inches, resulting in an additional 0.01 inch of tolerance for that joint, as shown in Figure 13.

**Figure 15**

*Deconstructed Fretboard Technical Drawing*

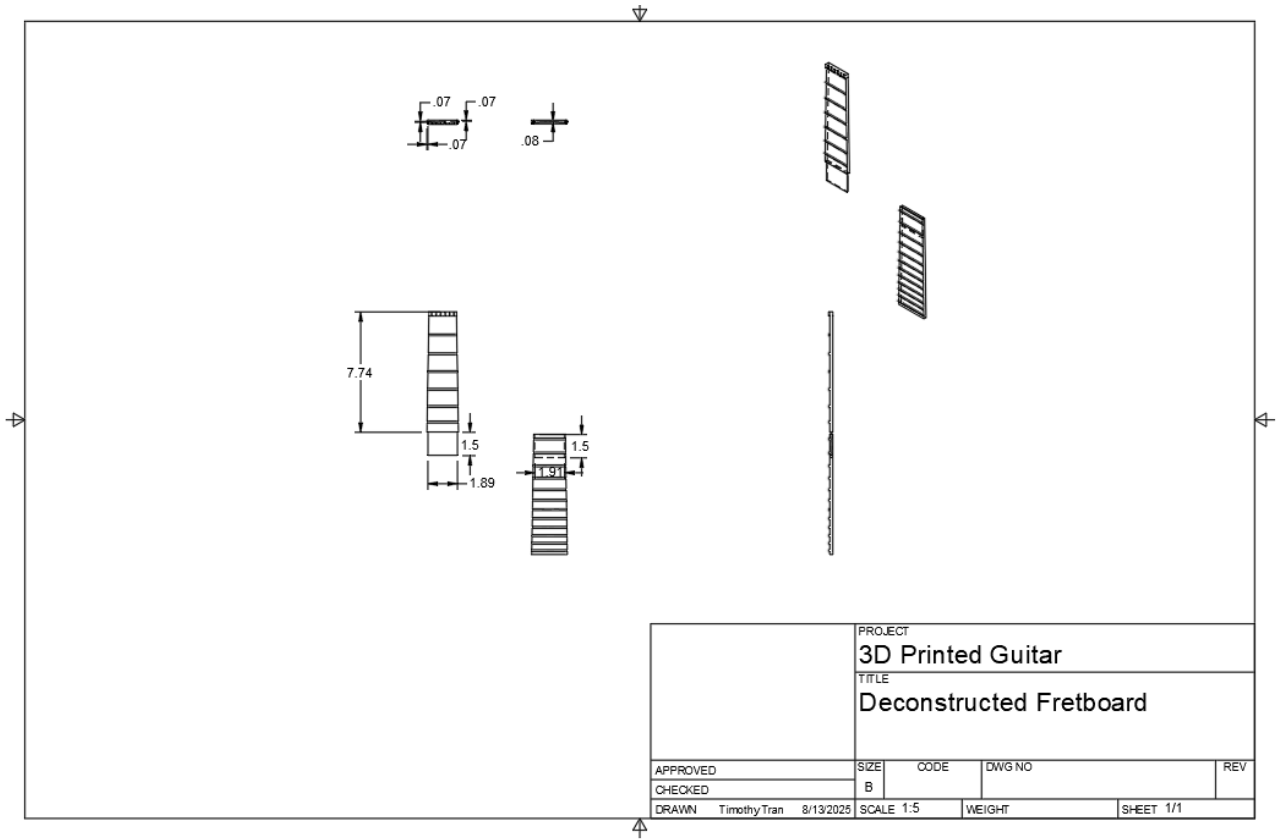

Figure 15 shows the fretboard bisection, with an added male and female head for connection. Initially, the fretboard connectors could not be press-fitted due to the smaller head size, and therefore, larger fillets of 0.05 inches were utilized.

Thinner components that are unsuitable for press-fitting were joined using a cyanoacrylate adhesive due to a lack of sufficient surface area to support a press-fit connection. Additionally, features such as the saddle must print facing upwards to ensure print quality, thus rendering connection heads on the bottom of such components inefficient due to the required orientation. Overhanging faces and component edges were integrated to promote better adhesion by increasing the points of contact for the adhesive. The top plate was divided as close to its center as possible, ensuring minimal stress when conjoining each of the constituent components, as seen in Figure 16.

**Figure 16**

*Deconstructed Top Plate Technical Drawing*

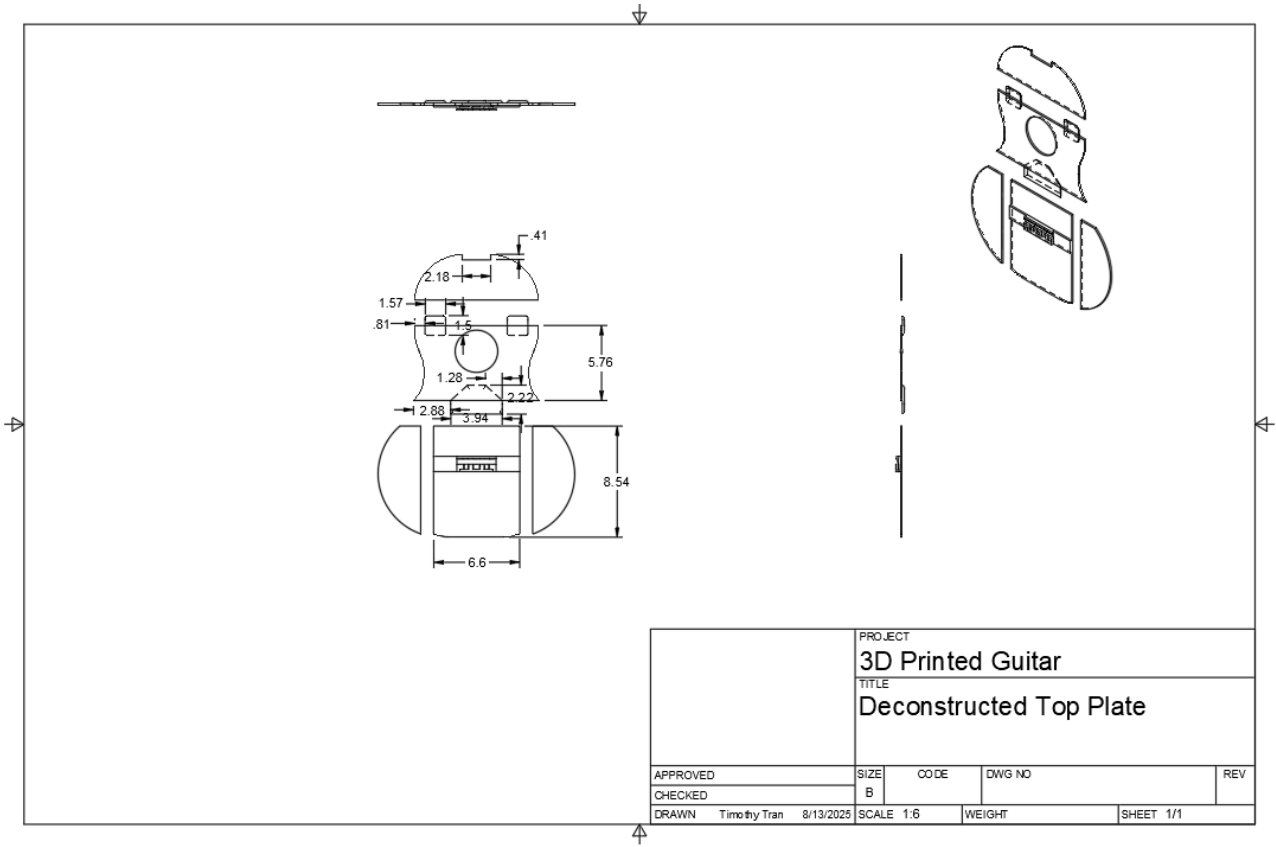

As shown in Figure 16, the front plate is divided into five components, utilizing overhangs as connection points for the cyanoacrylate adhesive. A cutout at the top of the plate is required to accommodate a reinforcement component, which was added between the neck and back plate to enhance stability through increased points of adhesion.

**Figure 17**

*Reinforcement Piece Technical Drawing*

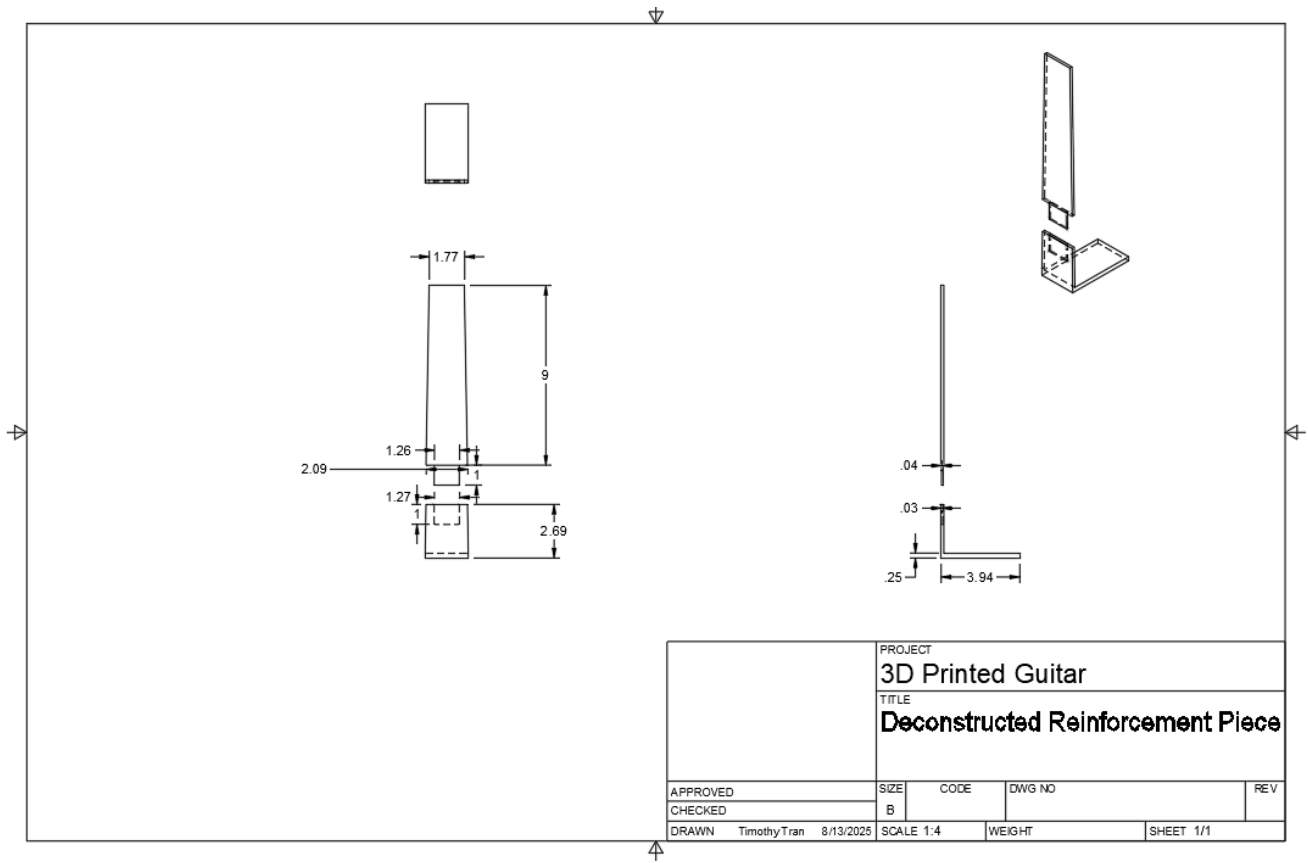

As shown in Figure 17, the reinforcement component features another press-fit connection. Additionally, clearances of 0.01 inch were applied at the base, preventing constricted assembly with the front plate cutout.

## Guitar Assembly and Analysis

Assembly is implemented using smaller parts, with tolerances adjusted as necessary. The body was assembled using the aforementioned 0.006-inch tolerance, starting with the neck base and the neck. The bottom plate utilized a press-fit tolerance consistently for a tight and precise fit. The reinforcement piece was printed and attached to the neck and bottom plate, creating the base of the guitar. Following this, the top plate was printed and connected using a cyanoacrylate adhesive, with the fretboard being placed on top. The nylon strings and tuning head were installed. The stress from tuning each string did not initially appear to create any visible damage on the guitar; however, during the time spent conducting research, slight cracks developed at the tuning head due to structural flex, potentially caused by the combination of summer heat and string tension. As shown in Figure 19, the constructed guitar retains its shape and structure with all of its requisite parts properly installed.

**Figure 19**

*Completed 3D Printed Guitar Prototype*

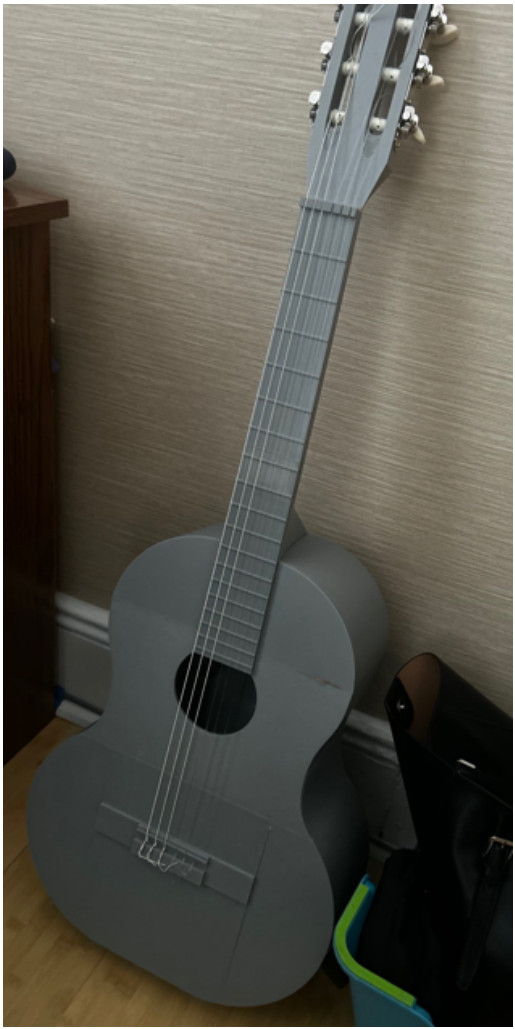

During tuning, a raised section at the fretboard connection joint pushed the frets upwards into the strings, muting their resonance. This issue was corrected by sanding down the raised point. The frequencies of each string were found using Audacity software, producing graphs that could be compared with the reference values (Mazzoni, 2000).

**Figure 20**

*Prototype String 1 Frequency Graph*

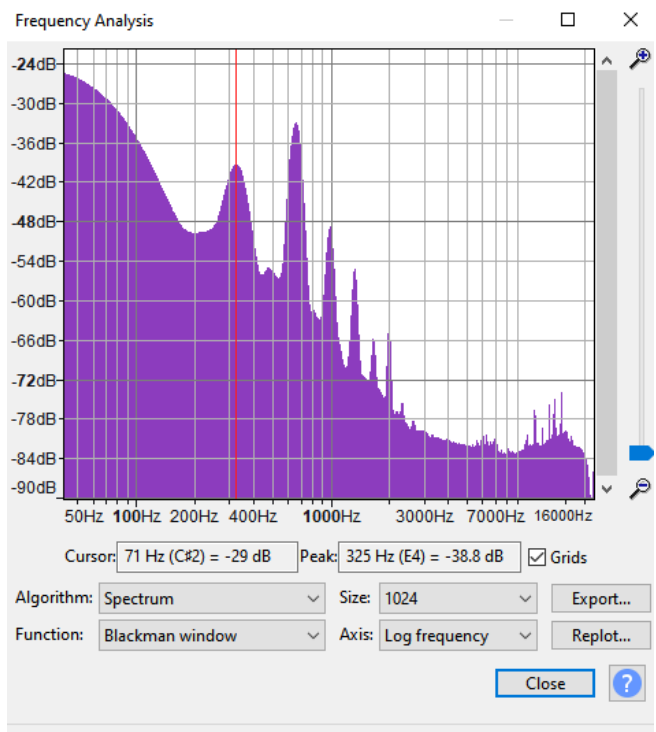

As shown in the graph in Figure 20, the first peak of the frequency is at the E4 note, with a frequency of 325 hertz, aligning closely with the standard frequency of the first string, E4, at 329.63 hertz.

**Figure 21**

*Prototype String 2 Frequency Graph*

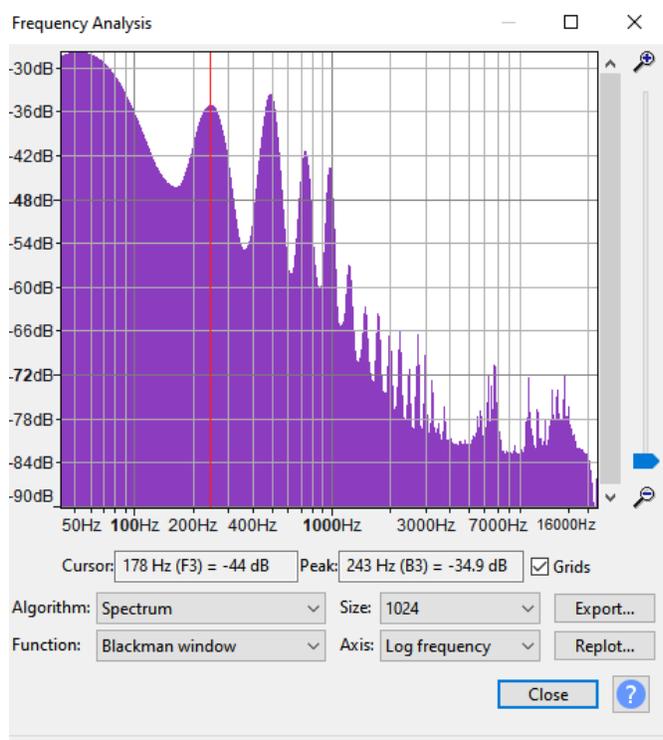

In Figure 21, the first peak at B3 note with a frequency of 243 hertz is observed, which is consistent with the standard B3 having a frequency of 246.94 hertz.

**Figure 22**

*Prototype String 3 Frequency Graph*

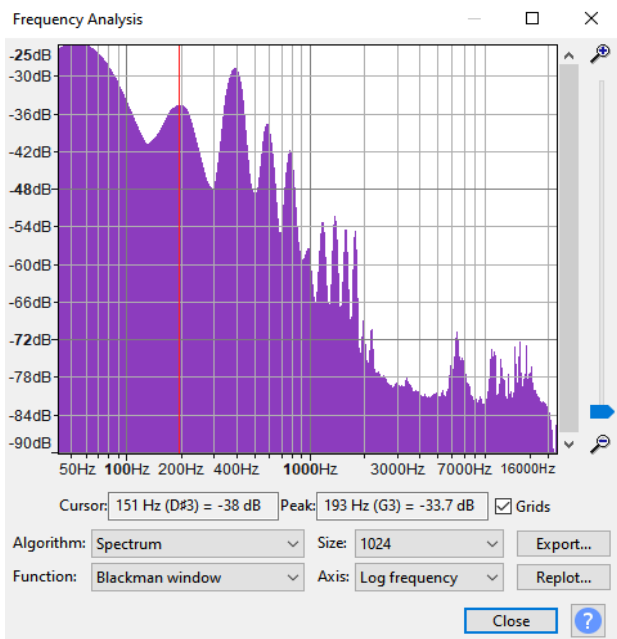

As seen in Figure 22, the initial peak is a note G3 with a frequency of 193 hertz, which is close to the standard G3 with a frequency of 196 hertz. The first three strings at the high end are significantly thinner than the bottom three, with the below graphs indicating a difference between the prototype printed guitar and traditional guitars.

**Figure 23**

*Prototype String 4 Frequency Graph*

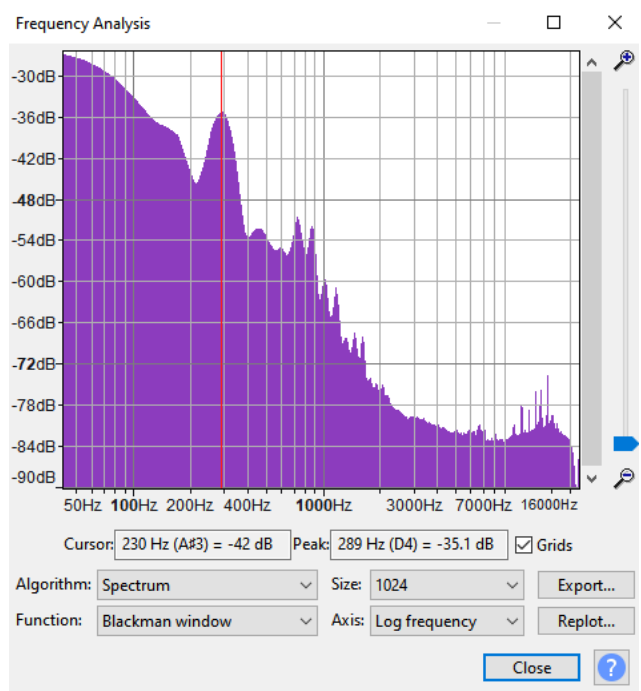

The first peak in Figure 23 is a D4 note with a frequency of 289 hertz. Though the note is very close to the standard D3, the frequency is nearly double compared to the 146.83 hertz standard. This may be due to the difference in sound projection in plastic materials, or it could be due to the string product chosen for this prototype.

**Figure 24**

*Prototype String 5 Frequency Graph*

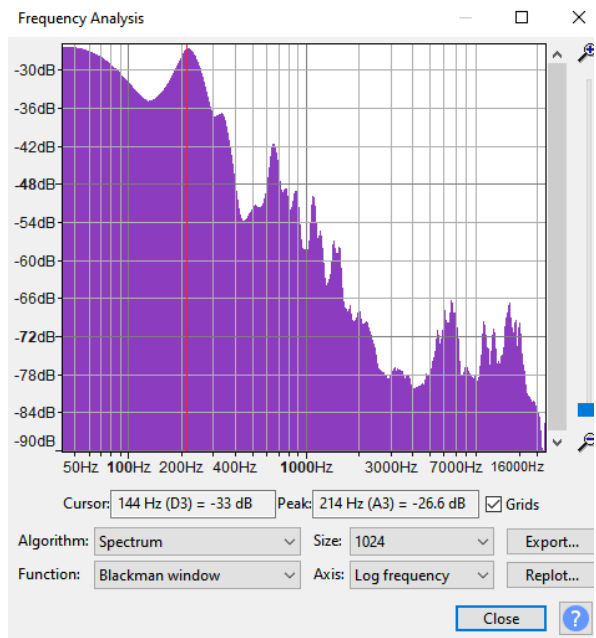

In Figure 24, the initial peak is an A3 note with a frequency of 214 hertz. The note is close to the standard of A2, but similar to the model in Figure 23, the frequency is nearly double compared to the projected 110 hertz.

**Figure 25**

*Prototype String 6 Frequency Graph*

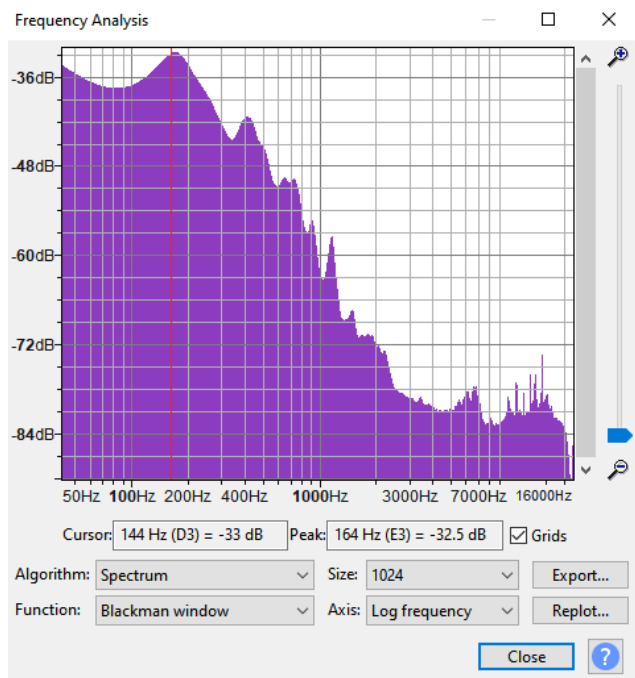

In Figure 25, the first peak is an E3 note with a frequency of 164 hertz. Similar to strings five and four, this note is accurate to the projected E2 but has nearly double the standard frequency of 82.41 hertz.

Tuning was relatively accurate in reaching the proper notes for the open strings with only some slight variations. Regarding the frequencies of each string, the first, second, and third strings were very similar to the projected standard, with an average deviation of nearly three hertz. However, looking at the fourth, fifth, and sixth strings, frequencies were almost doubled for each of them. This could be due to the PLA material properties or the particular string product chosen, thus creating differences in the prototype. Alternative printing plastics may be explored in the future to find a superior frequency match. Still, despite the foregoing, the tonality of the printed PLA guitar is nearly identical to that of a traditional guitar with no issues in tuning or sound projection.

**Social Impact**

The foregoing research on 3D printing acoustic guitar manufacturing has demonstrated that it is possible to engineer accessible musical instruments for everyone. The primary method examined herein, utilizing press-fit tolerancing with Polylactic Acid (PLA) components supplemented by minimal adhesive use, offers distinct advantages over the alternative approach of traditional wooden construction.

**Economic**

From an economic standpoint, printed guitars significantly reduce production costs. PLA is inexpensive, easily sourced, and requires no luthier training to use, thus allowing for easier production. Traditional wooden guitars require skilled labor, costly tonewoods, and precision tooling, where even small mistakes could ruin the guitar. 3D-printed instruments offer an alternative entry point that encourages musicians, regardless of their economic status.

**Technical**

From a technical standpoint, the press-fit method enhances the modularity of guitar production. The easy swapping of interchangeable components allows for quick repairs of a guitar should damage occur. Compared to traditional guitars, any damage to the wooden body could jeopardize the guitar's longevity, potentially requiring a full breakdown and reconstruction to repair cracks or tears. The innovative method of production described herein enables quick repairs and potential customization, allowing musicians to try new body shapes and observe how these changes impact the sound and enjoyment of the instrument.

**Educational**

An educational impact is primarily evident through positive effects on music curriculums and community involvement. Affordable and easily produced, 3D printed guitars provide an easy entry point for anyone interested in guitars or music. This accessibility fosters musical expression by enabling more individuals to participate in playing music, and the inherent nature of 3D printing and CAD allows people to share their ideas and models, encouraging collaboration among musicians and engineers alike.

**Environmental**

From an environmental standpoint, printed guitars have a surprisingly low impact on our world. PLA has a myriad of advantages over other plastics such as "... emit[ting] less toxic fumes than oil based plastics", being "... made from renewable raw materials", and being compostable (Column, 2020, para. 12). Although still impacting the environment, the drawbacks of using PLA in printed production are negligible compared to the harm of most other plastics.

Additionally, considering that "... species like mahogany, ebony, and rosewood have become endangered" (Rifkin, 2007, p. 1), PLA could be a great alternative to preserve these trees and look at other material sources for guitar making.

 The use of PLA in 3D printed guitars offers numerous benefits, including accessibility and ease of modification. Since these designs are digital, inexpensive, and easy to produce, 3D-printed guitars could potentially encourage more people to try guitar playing, as they do not require as much of a financial commitment. PLA could be the answer to new musical instrument production that benefits both people and the environment, especially given the dwindling resources of quality tonewoods in our modern world.

## Conclusion

 The foregoing research and prototype development confirm that additive manufacturing can produce a guitar capable of replicating the tonal range and accuracy of a standard wooden instrument. The use of press-fit tolerances, combined with cyanoacrylate adhesive, proved to be an effective assembly strategy, facilitating both rapid construction and convenient disassembly, if needed. The use of PLA enabled the fabrication of a fully functional instrument at a significantly reduced cost compared to the price of traditional tonewoods and the expertise of a luthier. Although future work should focus on refining the string frequency values to more closely align with those of conventional tonewoods, the current prototype demonstrates satisfactory tonal consistency, mechanical integrity, and overall performance.

 The implications of this work extend beyond its technical points into the broader social domain of music. By utilizing accessible, cost-effective materials through PLA, 3D-printed guitars can help reduce financial barriers for individuals seeking to explore music, particularly those in disadvantaged communities. This approach also promotes community through its potential customization and modeling in the CAD world. Along with the sustainability and repairability of printing a guitar in modular segments, this method of instrument production aligns with modern priorities in both consumer products and environmental responsibility. Moreover, it demonstrates how engineering innovation can work in tandem with artistic expression to offer a pathway towards inclusive musical instrument production for the future.